\documentclass[aps,prl,twocolumn,superscriptaddress,showpacs]{revtex4}
\usepackage{graphicx}
\usepackage{longtable}
\usepackage{array}
\usepackage{braket}
\usepackage{xfrac}
\newcolumntype{L}{>{\centering\arraybackslash}m{2cm}}
\newcolumntype{R}{>{\centering\arraybackslash}m{1.5cm}}
\newcolumntype{K}{>{\centering\arraybackslash}m{1.3cm}}
\setcitestyle{square}
\usepackage{gensymb} 

\usepackage{xcolor}

\def \ETO{Er$_2$Ti$_2$O$_7$}

\begin{document}

\title{Experimental evidence for field induced emergent clock anisotropies\\ in the XY pyrochlore \ETO}

\author{J.~Gaudet}
\email{gaudej@mcmaster.ca}
\affiliation{Department of Physics and Astronomy, McMaster University, Hamilton, Ontario L8S 4M1, Canada}

 \author{A.~M.~Hallas}
\affiliation{Department of Physics and Astronomy, McMaster University, Hamilton, Ontario L8S 4M1, Canada}
 
\author{J.~Thibault} 
\affiliation{Department of Physics and Astronomy, McMaster University, Hamilton, Ontario L8S 4M1, Canada}

\author{N.~P.~Butch}
\affiliation{NIST Center for Neutron Research, National Institute of Standards and Technology, Gaithersburg, Maryland 20899, USA}

\author{H.~A.~Dabkowska}
\affiliation{Brockhouse Institute for Materials Research, Hamilton, ON L8S 4M1 Canada}

\author{B.~D.~Gaulin}
\affiliation{Department of Physics and Astronomy, McMaster University, Hamilton, Ontario L8S 4M1, Canada}
\affiliation{Brockhouse Institute for Materials Research, Hamilton, ON L8S 4M1 Canada}
\affiliation{Canadian Institute for Materials Research, 180 Dundas Street West, Toronto, Ontario M5G 1Z8, Canada}

\date{\today}

\begin{abstract}

The XY pyrochlore antiferromagnet \ETO~exhibits a rare case of $Z_6$ discrete symmetry breaking in its $\psi_2$ magnetic ground state. Despite being well-studied theoretically, systems with high discrete symmetry breakings are uncommon in nature and, thus, \ETO~provides an experimental playground for the study of broken $Z_n$ symmetry, for $n>2$. A recent theoretical work examined the effect of a magnetic field on a pyrochlore lattice with broken $Z_6$ symmetry and applied it to \ETO. This study predicted multiple domain transitions depending on the crystallographic orientation of the magnetic field, inducing rich and controllable magnetothermodynamic behavior. In this work, we present neutron scattering measurements on \ETO~with a magnetic field applied along the [001] and [111] directions, and provide the first experimental observation of these exotic domain transitions. In a [001] field, we observe a $\psi_2$ to $\psi_3$ transition at a critical field of 0.18~$\pm$~0.05~T. We are thus able to extend the concept of the spin-flop transition, which has long been observed in Ising systems, to higher discrete $Z_n$ symmetries. In a [111] field, we observe a series of domain-based phase transitions for fields of 0.15~$\pm$~0.03~T and 0.40~$\pm$~0.03~T. We show that these field-induced transitions are consistent with the emergence of two-fold, three-fold and possibly six-fold Zeeman terms. Considering all the possible $\psi_2$ and $\psi_3$ domains, these Zeeman terms can be mapped onto an analog clock - exemplifying a literal clock anisotropy. Lastly, our quantitative analysis of the [001] domain transition in \ETO~is consistent with order-by-disorder as the dominant ground state selection mechanism.

\end{abstract}


\maketitle


\section{1. INTRODUCTION}

\begin{figure}[tbp]
\linespread{1}
\par
\includegraphics[width=3.3in]{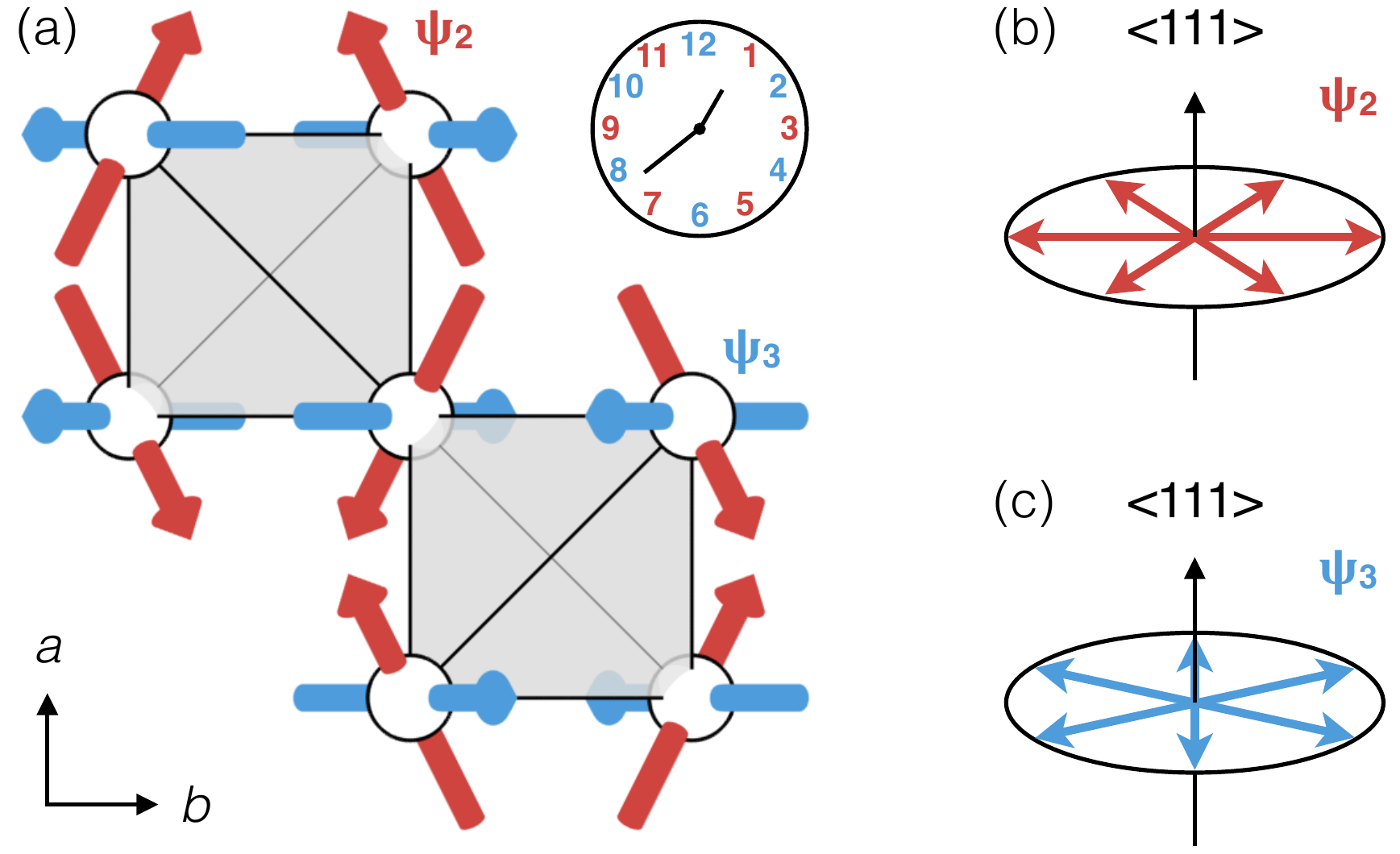}
\par
\caption{(a) The $k=0$, $\Gamma_5$ magnetic structure on the pyrochlore ($Fd\bar{3}m$) lattice, which is constructed by the $\psi_2$ (red) and $\psi_3$ (blue) basis vectors. Within the local XY plane, six specific spin orientations are allowed by (b) $\psi_2$ and six interleaving angles are allowed by (c) $\psi_3$. The ensemble of the full $\psi_2$ and $\psi_3$ states mimic a literal clock and can be used to represent the different anisotropic Zeeman terms via the selection of different hours (angles) on the clock.}
\label{struc}
\end{figure}

The pyrochlore lattice is a face centered cubic structure with a basis of corner-sharing tetrahedra. In the case of the pyrochlore \ETO, the spins residing on this network of tetrahedra are known to have a $k=0$, $\Gamma_5$ magnetic structure, for which all the spins lie in the plane perpendicular to the local $<$111$>$ axis~\cite{PhysRevB.68.020401}. A representation of this $\Gamma_5$ structure is shown in Fig.~\ref{struc}(a) with its associated basis vectors $\psi_2$ and $\psi_3$. The linear combination of these two basis vectors can generate any spin orientation spanning the local XY plane, which is the entire U(1) manifold. An appropriate model Hamiltonian that includes anisotropic exchange and dipolar interactions, with experimentally determined exchange parameters for \ETO, has its energy minimized by the U(1) manifold, which is degenerate at the mean field level~\cite{PhysRevLett.109.167201,PhysRevLett.109.077204}. However, in the real material, this degeneracy is lifted when the Er$^{3+}$ moments order antiferromagnetically into a pure $\psi_2$ state below $T_N=1.2$~K~\cite{Poole2007}. The mechanism responsible for this degeneracy breaking in \ETO~has attracted much attention, as it could be the first demonstration of ground state selection via order-by-disorder~\cite{Villain1980,Shender}. Indeed, it is widely believed that instead of ordering via energetic selection, thermal and quantum fluctuations drive this system to an entropically favorable magnetically ordered state~\cite{PhysRevLett.109.167201,PhysRevB.68.020401,Wong2013,PhysRevLett.109.077204,PhysRevB.88.220404,Javanparast2015}. This scenario is not yet definitive, however, as a competing theory has been proposed that might explain the ground-state selection via energetic selection~\cite{McClarty2009,PhysRevB.90.060410,Rau2015}. Regardless, \ETO~remains the most promising candidate for ground state selection via the order-by-disorder mechanism.\ 
        
Another interesting aspect of the phase transition in \ETO~is that it represents a rare case of higher order discrete symmetry breaking. Indeed, time inversion and rotational symmetry allow six distinct spin orientations within the $\psi_2$ state, \emph{ie.} a $Z_6$ symmetry breaking. The Ising model, with $Z_2$ symmetry breaking, is know{}n to capture the salient physics of many magnetic materials and a wealth of other physical systems~\cite{Chaikin2000,Schneidman2006}. Although many theoretical models for systems with higher $Z_n$ symmetry breaking have been proposed, there are in fact few experimental realizations of higher discrete symmetry~\cite{Pott2,Pott} Thus, \ETO~provides a rare and possibly unique opportunity to investigate the properties of such discrete symmetry breaking. Recently, Maryasin \emph{et al.}~\cite{Maryasin2016} developed a theor,y for the e,ffect of a magnetic field on the properties of a $Z_6$ pyrochlore magnet and predicted rich and exotic domain effects to occur in \ETO~due to the emergence of two-fold, three-fold and six-fold anisotropic Zeeman terms. Th,e degeneracy as well as the local XY angle of the states minimized by each of these Zeeman terms can be mapped out in analogy to a conventional clock, where the twelve hours of the clock are represented by the six $\psi_2$ and the six $\psi_3$ states (Fig.~\ref{struc}(b,c)).\ 

In this paper, we use time-of-flight neutron scattering on \ETO~with a magnetic field applied along high symmetry cubic directions to provide the first experimental evidence of these predicted domain effects. We find a host of low field domain selections and reorientations, which henceforth we will collectively refer to as ``domain transitions'', not to be confused with a change of representation manifold (U(1) or $\Gamma_5$) which would be a thermodynamic phase transition. Indeed, we find that for a field applied along the [001] direction, \ETO~exhibits a clear $\psi_2$ to $\psi_3$ transition at a critical field of 0.18~$\pm$~0.05~T. This domain transition can be seen as the $Z_n$ generalization of the spin-flop transition that occurs in Ising $Z_2$ systems. Our neutron scattering results also indicate possible domain transitions at 0.15~$\pm$~0.03~T and 0.40~$\pm$~0.03~T in a [111] magnetic field. We provide a complete description of the domains transitions that occurs in \ETO~in such fields and show that our observations are consistent with the predicted emergent Zeeman two-fold, three-fold and possibly six-fold clock terms.\

\section{II. EXPERIMENTAL DETAILS}

A large single crystal of \ETO~was grown in a floating zone image furnace in 3 atm.~of air and with a growth rate of 7~mm~h$^{-1}$. This method of crystal growth is well-established for the rare earth titanate pyrochlores~\cite{Li201396,balakrishnan1998single,Dabkowska2010}. This crystal was cut into two 2-3 gram segments, which were respectively aligned in the (H,K,0) and the (K+H,K$-$H,$-$2K) scattering planes using x-ray Laue diffraction. Time-of-flight neutron scattering measurements were performed using the Disc Chopper Spectrometer (DCS) at the NIST Center for Neutron Research~\cite{DCS}. An incident wavelength of 5 \AA~was employed, giving a maximum energy transfer of $\sim$2~meV and an energy resolution of 0.09~meV. All uncertainties correspond to one standard deviation. A magnetic field was applied perpendicular to the scattering plane. Thus, for the sample aligned in the (H,K,0) scattering plane, the field is applied along the [001] direction. For the sample aligned in the (K+H,K$-$H,$-$2K) scattering plane, the field is applied along the [111] direction. For both alignments, the (220) or (2-20) Bragg peak was observed only within the central bank of our detectors providing a 2$^\circ$ upper limit on the possible misalignment of our magnetic field. Scans with a total sample rotation of 35$\degree$~with 0.25$\degree$~steps were performed, centered on the (220) or (2-20) Bragg peak (these positions are symmetrically equivalent and will henceforth be referred to as (220)). Lastly, for completeness, we also present previously published measurements of \ETO~aligned in the (H,H,L) scattering plane, with a magnetic field applied along the [1-10] direction. This earlier experiment was performed on the same spectrometer (DCS at NIST) but with a different single crystal; the full experimental details can be found in Ref.~\cite{Ruff2008}.\

Typical elastic scattering maps of \ETO~for each of the three sample orientations above ($T = 8$~K or $2$~K) and below ($T = 60$~mK or $30$~mK) the Neel ordering transition are shown in Fig.~\ref{Qslice}. At high temperature, we observe a resolution limited Bragg peak that is purely structural in origin. At low temperature, passing into the magnetically ordered state, additional magnetic Bragg and diffuse scattering can be observed at the (220) position for all sample orientations. In the subsequent analysis, elastic cuts have been extracted for each data set with varying magnetic field, as indicated by the white dashed lines in Fig.~\ref{Qslice}(a,c,e). Those cuts have been obtained by integrating the respective data sets in energy from $-0.1$ to 0.1 meV with an additional integration (i) from $-0.3$ to 0.3 in the [0,0,L] direction for the [1-10] sample, (ii) from 1.8 to 2.2 in the [H00] direction for the [001] sample, and (iii) from 0.8 to 1.2 in the [H',0,-H'] direction for the [111] sample (Fig.~\ref{Elascuts001}(a,c,e)). The inelastic spectra, which are shown for each field direction in Fig.~\ref{Eslice}, are extracted by using the same directional integrations, but without the integration in energy. Integrations of the total inelastic signal about (220) are presented in Fig.~\ref{Ecuts}, where the area of integration corresponds to the white dashed boxes in Fig.~\ref{Qslice}(b,d,f). Those spectra are obtained using the same directional integrations as above, but with an additional integration (i) from 1.7 to 2.3 in [H,H,0] for the [1-10] sample, (ii) from 1.8 to 2.2 in [0K0] for the [001] sample, and (iii) from 0.8 to 1.2 in [K',-2K',K'] for the [111] sample.

\section{III. Results}

\subsection{A. Magnetic Field Dependence of the Elastic Scattering at (220)}

\begin{figure}[tbp]
\linespread{1}
\par
\includegraphics[width=3.3in]{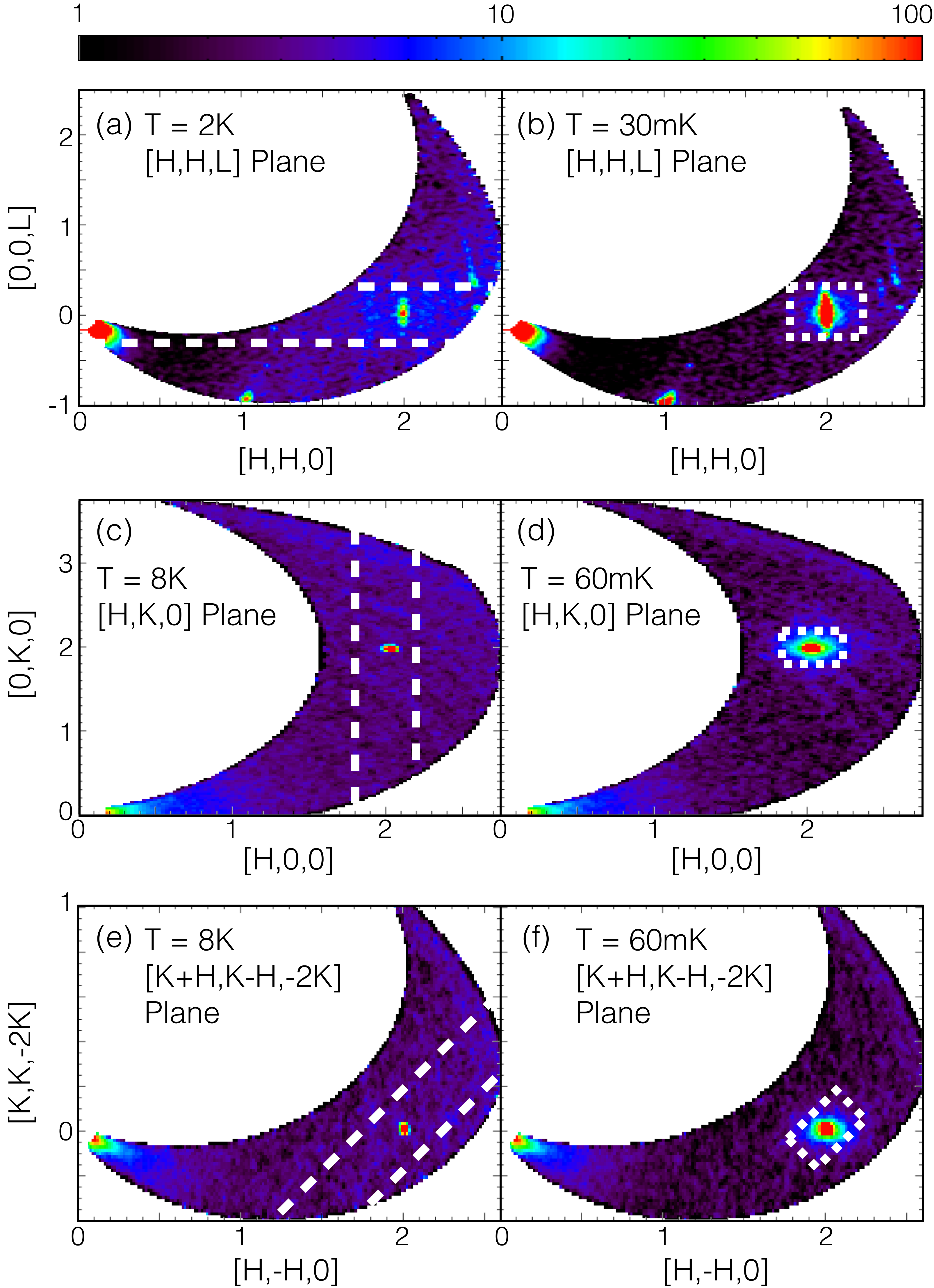}
\par
\caption{The elastic scattering of Er$_2$Ti$_2$O$_7$ above and below the Neel ordering transitions for crystals aligned in the (a,b) (H,H,L) plane, (c,d) (H,K,0) plane, and (e,f) (K+H,K$-$H,$-$2K) plane. At high temperature, the intensity of the (220) Bragg peak is purely structural, while at low temperature a magnetic Bragg peak also forms on the (220) position. Each of these data sets has been integrated in energy from $-0.1$ to 0.1 meV, which is approximately the resolution of elastic scattering in this experiment. Within the magnetically ordered state, this integration picks up a component of the magnetic inelastic scattering, giving the appearance of a significantly broadened peak. The dashed white lines indicate the areas of integration described in the text.}
\label{Qslice}
\end{figure}

Neutron scattering spectra of \ETO~were collected at very low temperature, below $100$~mK, with a magnetic field ranging from 0~T to 3~T applied along three crystallographic directions: [1-10], [001], and [111]. As can be seen by comparing the high temperature data sets (Fig.~\ref{Qslice}(a,c,e)) with the low temperature data sets (Fig.~\ref{Qslice}(b,d,f), there is considerable magnetic diffuse scattering at low temperature for all field directions. This diffuse scattering is far broader than a resolution limited Bragg peak typical of long range order. The origin of this diffuse scattering is an intense quasi Goldstone-mode, which softens towards (220)~\cite{Ruff2008,PhysRevLett.109.167201,PhysRevB.90.060410}. The quasi Goldstone-mode excitations have been previously measured in detail and are known to be gapped by 0.053~$\pm$0.006~meV~\cite{PhysRevLett.112.057201,PhysRevB.90.060410}. As the energy resolution of this experiment is 0.09~meV, we inevitably integrate over a portion of this low energy inelastic scattering when extracting the elastic component. Thus, the diffuse scattering observed in Fig.~\ref{Qslice}(b,d,f) originates from a partial integration of the quasi Goldstone-mode magnetic excitations.\

\begin{figure}[htbp]
\linespread{1}
\par
\includegraphics[width=3.3in]{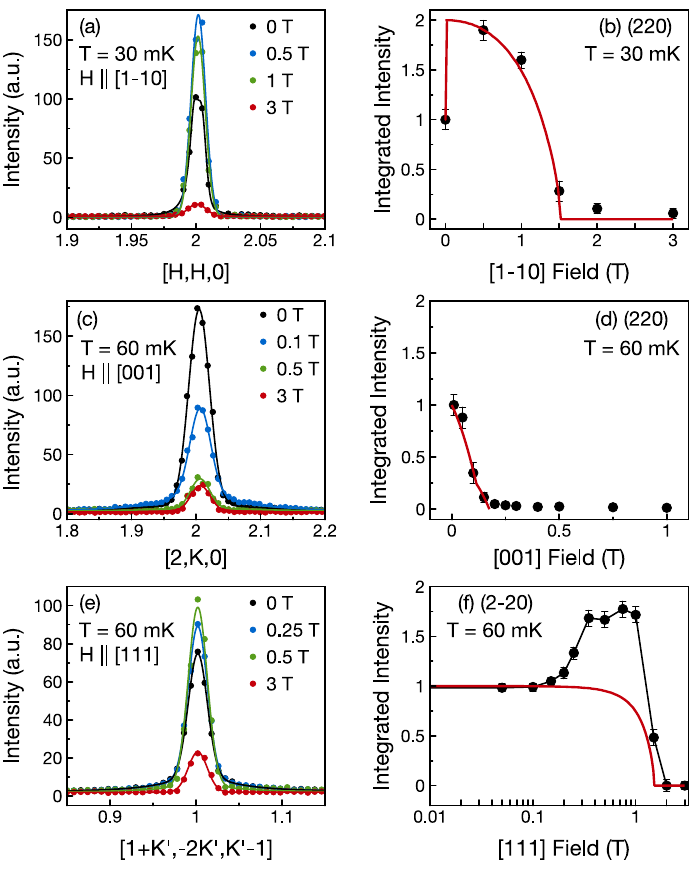}
\par
\caption{Representative selection of elastic cuts over the (220) Bragg peak in varying magnetic field strength, for fields applied along (a) the [1-10] direction, (c) the [001] direction, and (e) the [111] direction. The solid lines in these panels are the fits to the Bragg peak, which were used to extract the integrated intensity. The resultant field dependence of the magnetic elastic intensity at (220) with the field applied along the (b) [1-10], (d) [001], and (f) [111] direction, revealing multiple low field domain transitions in \ETO. The red curves in these panels corresponds to the theoretically predicted domain transitions~\cite{Maryasin2016}. Note that the (2-20) Bragg position is symmetrically equivalent to (220), as it is referred to in the main text.}
\label{Elascuts001}
\end{figure}

To investigate the field dependence of the elastic scattering, we performed integrations along the (220) Bragg peak for each data set at low temperature, with varying field strength and direction. A representative selection of these elastic scattering cuts are shown in Fig.~\ref{Elascuts001}(a,c,e). At low temperature, there is a significant increase in the intensity of the resolution-limited Bragg scattering, due to the long range magnetic order, and also, a diffuse contribution originating from the inelastic scattering as discussed above. These two contributions to the scattering at the magnetic Bragg position necessitate a two component fit~\cite{Gaudet2016EYTO}. These fits were performed with a Gaussian function, to account for the resolution limited elastic Bragg scattering, and a Lorentzian function, which captures the inelastic contribution to the scattering. The $Q$-width of the elastic Gaussian peak was fixed to the value determined from fitting the high temperature data set, for which the scattering at (220) is purely structural. The width of the Lorentzian function was allowed to freely vary. A sloping background was also used to account for the instrumental background. The solid lines seen in Fig.~\ref{Elascuts001}(a,c,e) are examples of typical fits obtained from following this procedure, giving excellent agreement with the measured data. The resulting integrated intensity for the elastic (Gaussian) component as a function of the [1-10], [001], and [111] applied magnetic field are shown in Fig.~\ref{Elascuts001}(b,d,f), respectively. The field dependence of the Lorentzian part of the scattering is presented in the Appendix A for all field directions.\

The elastic scattering dependence of the (220) magnetic Bragg peak in a [1-10] field is shown in Fig.~\ref{Elascuts001}(b). At low field, we observe an abrupt doubling of the intensity at (220). Above 0.5~T, the elastic scattering smoothly diminishes as a function of field and reaches zero intensity by 1.5~T. This smooth diminution of the (220) elastic scattering at high field corresponds to the transition towards the field polarized state~\cite{Maryasin2016,Bonville2013,Sosin2010,Ruff2008,Cao2010}. For the [001] field direction, the magnetic elastic intensity at (220) falls off precipitously under the application of a small field (Fig.~\ref{Elascuts001}(d)). Indeed, the elastic magnetic scattering at (220) reaches zero intensity in a field as small as 0.2~T. As the field is further increased, up to 3~T, the intensity remains zero. Finally, for the [111] field, the elastic scattering at (220) is unaffected up to 0.15~T (Fig.~\ref{Elascuts001}(f)). Between 0.15~T and 0.4~T, the elastic scattering abruptly increases, reaching an intensity that is 1.75 times larger than the elastic intensity at 0~T. The intensity then remains constant from 0.4~T to 1~T. Above 1~T, the intensity decreases, ultimately reaching zero intensity for fields larger than 1.5~T. This decrease of the elastic scattering at (220) above 1.5~T is, as before, concomitant with the phase transition towards the field polarized state~\cite{Bonville2013,Sosin2010}.\

\begin{figure*}[tbp]
\linespread{1}
\par
\includegraphics[width=7in]{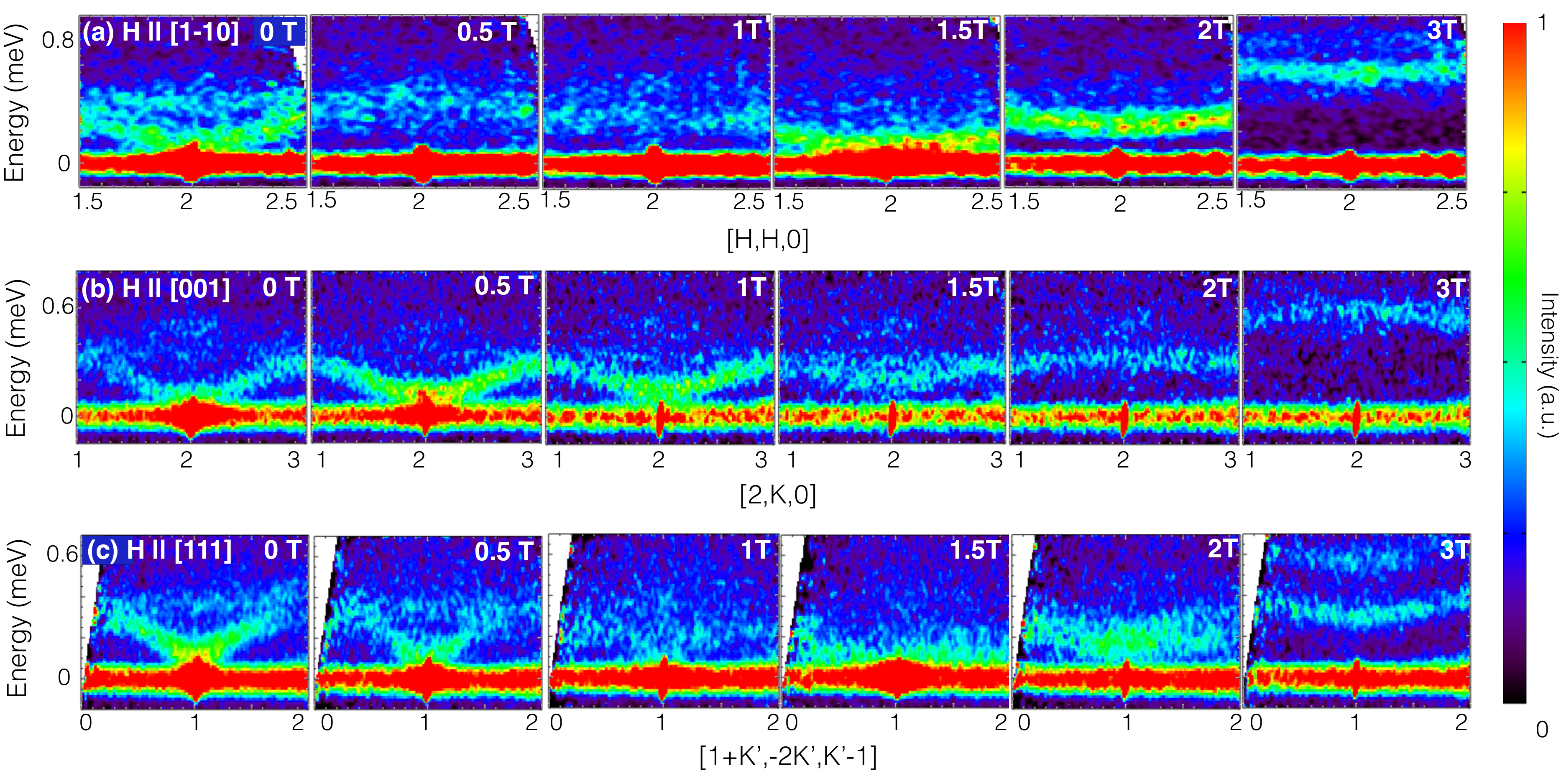}
\par
\caption{Magnetic field dependence of the energy slices of \ETO~centered on the (220) Bragg position. These energy slices are plotted along (a) the [H,H,0] direction for a [1-10] field, (b) the [0,K,0] direction for a [001] field, and (c) the [K',-2K',K'] direction for a [111] field. For the [1-10] field orientation, there is an immediate diminution of the quasi Goldstone-mode excitations at 0.5~T. For the [001] field orientation, the quasi-Goldstone mode excitations at (220) are intensified by the application of a field up to 1~T. In a [111] field, the quasi-Goldstone mode excitations have their intensity continuously decreased for fields ranging from 0.15 to 1~T. Above 1.5~T, the response of the spin wave spectra is due to the transition towards the field polarized state.}
\label{Eslice}
\end{figure*}

\subsection{B. Magnetic Field Dependence of the Inelastic Scattering at (220)}

\begin{figure*}[tbp]
\linespread{1}
\par
\includegraphics[width=6.5in]{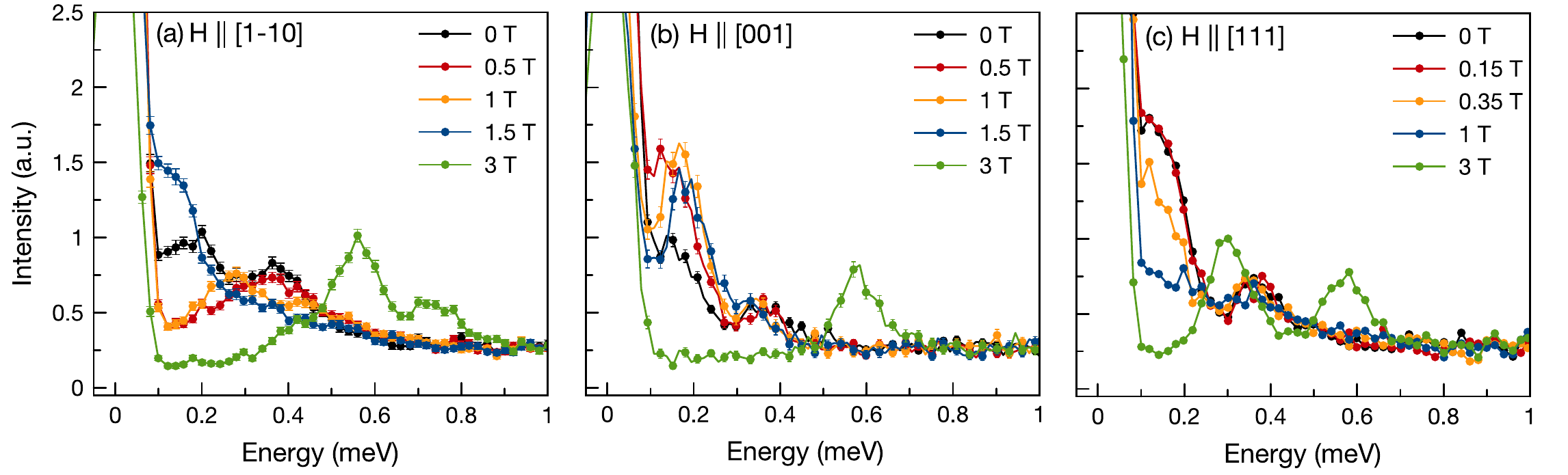}
\par
\caption{The inelastic intensity centered on (220) as a function of energy for varying (a) [1-10] magnetic fields, (b) [001] magnetic fields and (c) [111] magnetic fields. In a [1-10] field, the low energy scattering is completely suppressed upon the application of a 0.5~T field. In a [001] field, the inelastic intensity at low energies strongly increases for fields as small as 0.1~T. Further increasing the field shifts the spectral weight to higher energies. In a [111] field, the inelastic spectra is unaffected by fields up to 0.15~T. At larger [111] fields, the intensity at low energies continuously decreases. The integration performed in Q are given in the experimental methods section.}
\label{Ecuts}
\end{figure*}

Now, turning our attention to the inelastic scattering, we can first look at Fig.~\ref{Eslice}, which shows the spin wave spectra of \ETO~as a function of [1-10], [001], and [111] field. Integrations as a function of energy centered on the (220) Bragg peak for each field direction are also shown in Fig.~\ref{Ecuts}. The spin wave spectra at 0~T is dominated by low energy quasi Goldstone-mode excitations centered on (220). These low energy excitations have a linear dispersion and are maximally intense approaching the (220) Bragg position. Accordingly, in the 0~T energy cuts of Fig.~\ref{Ecuts}, the integration over the quasi Goldstone-mode excitations produces the first inelastic feature, centered just below 0.2~meV. At slightly higher energy, 0.35~meV, we also observe weaker flat modes. The structure of the spin wave spectra in an applied field are further analyzed in, first, the low field regime and second, in the high field regime crossing into the field polarized state. The critical field of the transition to the field polarized state depends on the field orientation and is known, from bulk measurements, to be $1.5-1.7$~T~\cite{Bonville2013,Sosin2010}.   

The application of a small [1-10] field results in a complete absence of low energy scattering at 0.5~T, indicating the removal of the quasi Goldstone-mode excitations at (220) (Fig.~\ref{Eslice}(a) and \ref{Ecuts}(a)). However, the opposite scenario is observed upon the application of a weak [001] field. Indeed, an immediate increase in the scattering of the quasi Goldstone-mode excitations is observed. This enhancement of the scattering can be seen by comparing the 0~T and 0.5~T data sets in Fig.~\ref{Eslice}(b) and \ref{Ecuts}(b). For the [111] field direction, the quasi Goldstone-mode excitations shows no field dependence up to 0.15~T (Fig.~\ref{Eslice}(c) and \ref{Ecuts}(c)). Above 0.15~T, we observe a suppression of the quasi Goldstone-modes excitations, but without their complete removal, as in the case of a [1-10] field. In fact, while continuously decreasing up to 1~T, the intensity of these excitations remains finite in a [111] field.\   

Lastly, we can examine the changes in the spin wave spectra upon transitioning into the field polarized state, which is known to occur at $1.5-1.7$~T~\cite{Bonville2013,Sosin2010}. The evolution of the spin wave spectra when passing into the field polarized state gives similar behavior for [1-10] and [111] fields. Fig.~\ref{Eslice}(a) and (c) show that the spectral weight for these two field orientations softens towards the elastic line in a 1.5~T field. Upon further increasing the magnetic field to 3~T, this quasi-elastic scattering moves to higher energies, and forms weakly-dispersing spin wave modes. A qualitatively different behavior is observed in the case of a [001] field. Approaching the polarized state for a [001] field, coherent low energy excitations are still observed, but with an increasing spin wave gap at (220). In Fig.~\ref{Ecuts}(b), the opening of the spin wave gap is demonstrated by the shifting of the low energy feature upwards in energy. Upon the application of a [001] field greater than 1.5~T, the dispersion of the low energy quasi Goldstone-mode excitations smoothly evolve to a non-dispersive mode, as seen at 3~T. 

\section{IV. Discussion}

The magnetic ground state of \ETO~in zero magnetic field is well-established and corresponds to an equiprobable distribution of the six $\psi_2$ domains within $\Gamma_5$~\cite{Poole2007}. The elastic neutron scattering profile of this magnetic structure is characterized, in part, by an intense (220) magnetic Bragg peak~\cite{Ruff2008,PhysRevB.68.020401,McClarty2009}. Once a weak magnetic field is applied along any direction, the degeneracy of the six $\psi_2$ domains is lifted. However, it is important to re-emphasize that this does not correspond to a thermodynamic phase transition or a change of representation manifold. Rather, the spins remain constrained to the U(1) plane ($\Gamma_5$ manifold), as has been previously shown by heat capacity~\cite{Sosin2010,Ruff2008}, magnetization~\cite{Bonville2013}, and neutron scattering~\cite{Cao2010,Ruff2008}. Depending on the field orientation, this degeneracy breaking results in an increase or decrease of the (220) Bragg peak intensity due to domain effects. To understand these intensity changes, it is important to understand that the scattered intensity at (220) follows an $I\propto\cos^2(\theta)$ relationship, where $\theta$ is the local XY angle (Fig.~\ref{Calculation}). As neutron scattering is only sensitive to the component of the magnetization perpendicular to the direction of the scattering vector, a variation in the scattered intensity is observed due to the different orientations of the moments in each domain. Thus, by using the results of the calculation in Fig.~\ref{Calculation} and the fact that the magnetic state at 0~T is well known (six $\psi_2$ states), it is possible to deduce the reorientation of the domains that occur in a magnetic field by measuring the relative change of the (220) elastic intensity.\

\subsection{A. Domain Selection in a [1-10] Magnetic Field}

Before discussing the [001] and [111] field evolution of the (220) Bragg peak, we briefly review the well-established domain effects for a field applied along the [1-10] direction. For this field orientation, an increase of the scattering at (220) is observed for fields above 0.1~T (Fig.~\ref{Elascuts001}(b))~\cite{PhysRevB.68.020401,Ruff2008}. The origin of this intensity gain is well-understood: the application of a [1-10] field on \ETO~in its $\psi_2$ magnetic ground state induces a two-fold clock term that favors the $\psi_2$ states with XY angles of 0 and $\pi$~\cite{PhysRevLett.109.167201,Maryasin2016}. These two angles are highlighted by the dashed black circles in Fig.~\ref{Calculation}(a), and are the ones that maximize the scattered intensity of the (220) magnetic Bragg peak. These two angles give a factor two intensity increase to (220) from the average value for the six zero field $\psi_2$ states. These two domains selected by the Zeeman-clock term for a [1-10] field are also selected by the six-fold anisotropic term at 0~T. No further domain transitions are observed or predicted at low field. Beyond that, at high field, a continuous transition towards the field polarized state is observed at 1.5~T which is indicated by the smooth diminution of the (220) intensity (Fig.~\ref{Elascuts001}(b)). This diminution of the (220) intensity occurs due to canting effects that become non-negligible approaching the field polarized state. This canting effect introduces a non-zero spin component away from the XY local plane. Several studies have modeled this behavior~\cite{Cao2010,Ruff2008,McClarty2009}, but we reproduced via the red line in Fig.~\ref{Elascuts001}(b), the results using the method found in Maryasin et al.~\cite{Maryasin2016} which capture well the experimental data.\

\subsection{B. Domain Selection and Reorientation in a [001] Magnetic Field} 

As was the case for a [1-10] field, it has been predicted that in a [001] field, the Zeeman coupling will give rise to a two-fold clock term~\cite{Maryasin2016}. The states that are selected in a [001] field are rotated by $\pi/2$ with respect to the states favored by a [1-10] field. Thus, the Zeeman energy is fully minimized by the $\psi_3$ states with XY local angles of $\pi/2$ and $3\pi/2$. However, at very low fields, where the Zeeman energy is much smaller than the six-fold $\psi_2$ clock term that dominates at 0~T, it is not possible to select these two $\psi_3$ domains. Instead, the system compromises by selecting the nearest $\psi_2$ domains, those with XY local angles of $2\pi/6$, $4\pi/6$, $8\pi/6$ and $10\pi/6$ (See black dashed circles on Fig.~\ref{Calculation}(b)). Compared to 0~T, this new domain distribution should decrease the (220) intensity by a factor two. Experimentally, referring back to Fig.~\ref{Elascuts001}(b), we do indeed observe a clear decrease of the scattering at low field.\

The application of a field along [001] is inherently more interesting than a [1-10] field, as no $\psi_2$ state can fully minimize the Zeeman energy. Thus, there is competition between the emergent two-fold Zeeman term and the 0~T six-fold clock term. At higher fields, as the Zeeman energy begins to overwhelm the 0~T six-fold clock term, there will be a continuous rotation of the spins towards a pure $\psi_3$ state made up of $\pi/2$ and $3\pi/2$ domains, where the (220) intensity should decrease to zero. In our experiment, the intensity of (220) continuously and drastically decreases to reach zero intensity by 0.2~T (Fig.~\ref{Elascuts001}(b)). We associate this dramatic intensity loss with the above-described $\psi_2$ to $\psi_3$ transition, which is an XY spin-flop transition. As opposed to the $\psi_2$ state, all the spins within the $\psi_3$ states point perpendicular to the [001] direction (see Fig.~\ref{struc}(a)). This result generalizes the well known concept of spin-flop transitions seen in Ising system ($Z_2$) to systems with discrete symmetry breaking $Z_n$ with $n>2$.\  

The domain transition in a [001] field also opens an interesting line of inquiry, as the critical field for the $\psi_2$ to $\psi_3$ transition occurs when the Zeeman energy equals the energy of the 0~T six-fold clock term. Thus, we are able to provide an independent measurement of the 0~T six-fold clock term in \ETO, which is of course relevant to the zero field ground state selection, be it order-by-disorder and/or energetic selection involving virtual crystal field processes. Using analytical equations from Ref.~\cite{Maryasin2016}, which account for the field dependence of the XY local angle upon the application of a [001] field, it is possible to model the decrease of the elastic scattering, the result of which is shown by the red line in Fig.~\ref{Elascuts001}(d). This fit is optimized as a function of the critical field, and the value we obtained is $0.18\pm0.05$~T. Calculations using only the quantum order-by-disorder term have predicted a critical field of 0.2~T~\cite{Maryasin2016}, which is in excellent agreement with our experimental value. Naively, such a result can be interpreted as the $\psi_2$ selection in \ETO~being largely dominated by order-by-disorder effects, with virtual crystal-field excitations contributing relatively little to the $\psi_2$ selection. However, the critical field for the $\psi_2$ to $\psi_3$ spin-flop transition can only be indirectly modeled, introducing a degree of uncertainty. Further analysis, perhaps via numerical methods, may allow a more accurate comparison between the strength of the 0~T order-by-disorder six-fold clock term and the experimental data. Nonetheless, in principle, our measurement of the 0~T six-fold clock term should provide important information that would allow the ground state selection in \ETO~to be definitively understood.\

\subsection{C. Domain Selection and Reorientation in a [111] Magnetic Field} 

The Zeeman coupling for a field along the [111] direction is predicted to induce a combination of a three-fold and six-fold clock terms and, hence, the domain behavior of \ETO~ in a [111] field is expected to be rich~\cite{Maryasin2016}. The three-fold clock term selects the $\psi_2$ states with angles corresponding to $\pm\pi/3$ and $\pi$. However, the six-fold clock term is not the same as the one that selects the 0~T $\psi_2$ states. Instead, this [111] six-fold Zeeman clock term selects states which are rotated with respect to both the $\psi_2$ and $\psi_3$ states. This term then competes with both the three-fold Zeeman clock term and also against the 0~T six-fold clock term that favors the $\psi_2$ states. 

\begin{figure}[tbp]
\linespread{1}
\par
\includegraphics[width=3.3in]{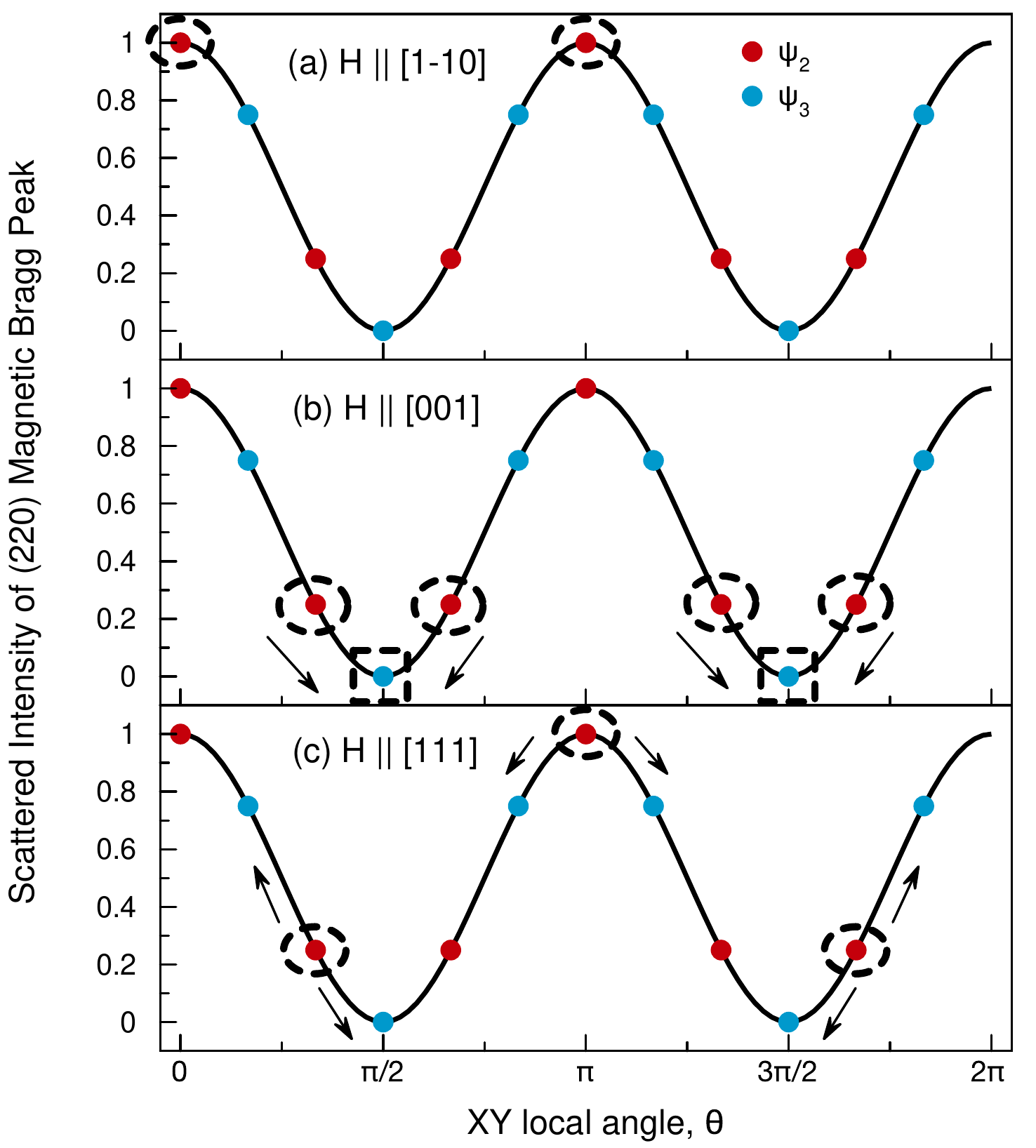}
\par
\caption{The scattered intensity at the (220) Bragg position in \ETO~as a function of the angle, $\theta$, in the XY local plane. The red dots represent the six $\psi_2$ domains and the blue dots represent the six $\psi_3$ domains. In (a) the two $\psi_2$ domains circled in black are the ones selected when a magnetic field of order 0.1~T is applied along the [1-10] direction. In (b) the four domains circled in black are the four $\psi_2$ states immediately selected in a [001] field. These states then cant towards the two $\psi_3$ states indicated by the black squares, which are selected by a 0.18~$\pm$~0.05~T [001] magnetic field. In (c) the three $\psi_2$ domains circled in black are the ones selected for magnetic fields up to 0.15~T applied along the [111] direction. At higher fields, these domains are predicted to split by an angle, $\theta$, as indicated by the arrows. Our measurement in a [111] field measured the (2-20) Bragg peak, which is symmetrically equivalent to (220) and gives the identical result as shown in panel (c)}
\label{Calculation}
\end{figure}

At low field, it is predicted that the domain selection behavior should be dominated by the combination of both the three-fold Zeeman clock term and the 0~T six-fold clock term. First, at very low fields, the three domains with an angle of $\pm\pi/3$ and $\pi$ should be selected (see black circles on Fig.~\ref{Calculation}(c)), which would result in no change in the (220) intensity. Experimentally, we refer back to our measurement shown in Fig.~\ref{Elascuts001}(f), which shows that the elastic scattering is constant up to 0.15~T. This is then consistent with the theoretical prediction that the three indicated $\psi_2$ states are selected for small [111] fields. Moreover, our data is in good agreement with the theoretical prediction that a weak [111] field induces an emergent three-fold clock term.\   

Upon further increasing the [111] field, it is predicted that \ETO~should experience two additional domain transitions prior to entering its classical field polarized state. The origin of these transitions is the emergent six-fold Zeeman clock term. The first of these transitions is theoretically predicted to occur at $H_{c1}=0.16$~T and results in the formation of six new domains. These six new domains are related to the previously selected $\pm\pi/3$ and $\pi$ domains, but with a splitting angle of $\pm\theta$ (see black arrows in Fig.~\ref{Calculation}(c)). This splitting angle, $\theta$, is predicted to have the same value for all domains and should increase as a function of the applied field. At still higher fields, $H_{c2}=0.4$~T, a second transition decreases the splitting angle back to zero, returning the system to a state with $\pm\pi/3$ and $\pi$ domains. These transitions would be difficult to verify using unpolarized neutron scattering, as the domain splitting by an angle, $\theta$ would have zero net effect on the intensity of (220). Thus, for a [111] field, the (220) Bragg peak is predicted to have no changes to its intensity up until the transition to the field polarized state. The predicted (220) behavior calculated by Maryasin et al.~\cite{Maryasin2016} is indicated by the red line in Fig.~\ref{Elascuts001}(f) and it is apparent that our experimental observations are not fully consistent with the predicted scenario, as the intensity of (220) is observed to substantially vary above 0.15~$\pm$~0.03~T. We note, however, the predicted critical fields for these transitions, $H_{c1}=0.16$~T and $H_{c2}=0.4$~T do appear to be meaningful in \ETO: the first, $H_{c1}$, corresponds with the observed increase in elastic intensity at (220) by a factor of 1.75 and the second, $H_{c2}$, corresponds quite well with the field at which the intensity flattens, 0.4~$\pm$~0.03~T.

To account for the observed intensity gain of (220) above 0.15~T, we propose a scenario of non-equiprobable domain distribution or inequivalent splitting angles (Fig.~\ref{Calculation}(c)). Such scenarios would explain the increase in the elastic scattering at (220), but would require an additional Zeeman clock term that would favors the $\pi$ domain over the $\pm\pi/3$ domains. In the theoretical work of Maryasin \emph{et al.}, no such term is predicted for a perfectly aligned sample~\cite{Maryasin2016}. One possible origin of such a Zeeman clock term is a slight misalignment of the field along the [111] direction. While we can place an upper bound of 2$^{\circ}$ on the error in our alignment, it is worth noting that very small misalignments in a [111] field are known to be enhanced by demagnetization effects in spin ice~\cite{Morris2009}. However, in contrast to the spin ice case, \ETO~is a antiferromagnet where such demagnetization effects are naively expected to be far less important. Thus, the precise domain distributions and orientations occurring for fields between 0.15 and 0.4~T in a [111] field remain an open question at present.\  

\subsection{D. Effect of Domain Selection on the Quasi Goldstone-Mode Excitations}

Finally, it is interesting to comment on the low field behavior of the quasi Goldstone-mode excitations centered on (220). In a small [1-10] field, as we discussed in the Results section III.B., the quasi Goldstone-mode excitations disappear at this particular ordering wavevector. This is best observed by comparing the inelastic spectra of 0~T and 0.5~T in Fig.~\ref{Eslice}(a) and Fig.~\ref{Ecuts}(a). The disappearance of the quasi Goldstone-mode excitations is concomitant with the increase of the elastic scattering at (220) (Fig.~\ref{Elascuts001}(b)). The exact opposite behavior is observed for a field along the [001] direction, where we observe that the quasi Goldstone-mode excitations increase for small fields (Fig.~\ref{Eslice}(b) and Fig.~\ref{Ecuts}(b)), while the elastic scattering (Fig.~\ref{Elascuts001}(d)) decreases for the same field range. Thus, there is a clear trade off between the intensities of the elastic scattering and the quasi Goldstone-mode scattering at the (220) magnetic Bragg peak while transitioning into states with different XY local angles. To understand this observation, we suggest that the intensity of the quasi Goldstone-mode at (220) goes like $I\propto\sin^2(\theta)$. This can be understood by first pointing out that for magnetic neutron scattering, it is the component of the moment perpendicular to Q that couples to the neutron (\emph{ie.} in this case perpendicular to (220)). The domains for which the (220) elastic scattering is maximized have their spin directions maximally perpendicular to the (220) direction, and vice-versa. However, for low fields, the spins remain constrained to lie within the local XY plane. Thus, in the case of a [1-10] field, where the spins are maximally perpendicular to the scattering vector, the quasi Goldstone-mode excitations must necessarily have their intensity reduced, as the spins will become more parallel to Q and will become less visible to neutron scattering. The opposite is true for a [001] field, where the spins are maximally parallel to the scattering vector and must excite into a more perpendicular orientation. The effect of a field on the scattered intensity of the quasi Goldstone-mode are thus, another direct signature of the domains effects. Moreover, the field dependence of the inelastic and elastic scattering in \ETO~are consistent with each other.\\

\section{V. Conclusions}

We have performed comprehensive time-of-flight neutron scattering measurements on single crystals of \ETO~with a magnetic field applied along the [001] and [111] directions. For fields smaller than 1~T, the field induced effects can be attributed to domain selection. The zero field state of \ETO~assumes the $\psi_2$ antiferromagnetic structure, which is composed of six equally probable domains. For small fields applied along the [001] direction, we observe a dramatic decrease of the (220) magnetic Bragg peak intensity, which agrees very well with the predicted transition from the six $\psi_2$ domains to two $\psi_3$ domains. For the [111] field direction, we observe that the elastic scattering at (220) is independent of field for low fields, consistent with an emergent three-fold Zeeman clock term. Further increasing the field results in a large enhancement of the (220) Bragg peak intensity, inconsistent with the predicted domain selection scenario and hinting at an even richer phase behavior. Lastly, our experiment provides a measure of the zero field six-fold clock term strength, 0.18~$\pm$~0.05~T. This result should prove useful in establishing a complete understanding of the mechanism of ground state selection in \ETO.

The work presented here experimentally confirms the rich and exotic magnetothermodynamic behavior predicted to exist in the XY pyrochlore antiferromagnet \ETO, or any other pyrochlore magnet ordering into the $\Gamma_5$ manifold. We demonstrate that, depending on the field direction, a combination of two-, three-, and possibly six-fold Zeeman clock terms emerge and compete with the six-fold clock term present in zero field. We anticipate this work will help determine the ground state selection and field effects in other topical  $\Gamma_5$ pyrochlores, such as NaCaCo$_2$F$_7$~\cite{Ross2016}, Er$_2$Ge$_2$O$_7$~\cite{Dun2015} and Yb$_2$Ge$_2$O$_7$~\cite{Hallas2016,Dun2015}.

\begin{acknowledgments}
We thank M. Zhitomirsky and R. Moessner for very useful discussions and for providing us the theoretical curves of the field dependence of the (220) Bragg peak intensity shown in Fig.~\ref{Elascuts001}(b,f). We also wish to thank Yegor Vekhov for his technical support with the dilution fridge used in the neutron experiment. The neutron scattering data were reduced and analyzed using DAVE software package~\cite{Dave}. The NIST Center for Neutron Research is supported in part by the National Science Foundation under Agreement No. DMR-094472. Works at McMaster University was supported by the National Sciences and Engineering Research Council of Canada (NSERC).  
\end{acknowledgments}

\begin{appendix}
\section{Appendix A: Field effects on the Lorentzian contribution of the (220) magnetic Bragg peak}

\begin{figure}[tbp]
\linespread{1}
\par
\includegraphics[width=3.3in]{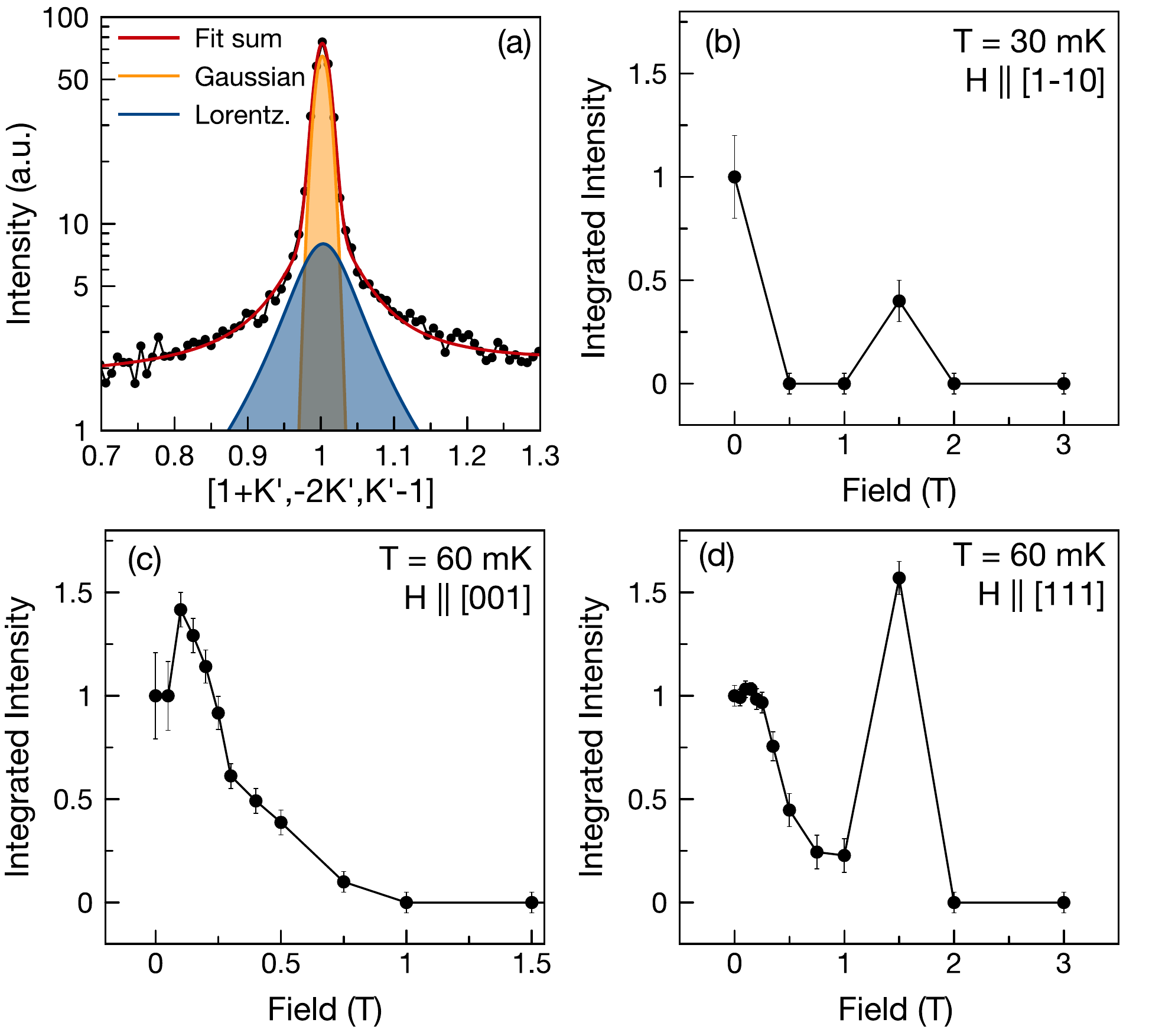}
\par
\caption{(a) A typical fit of the scattering at the (220) Bragg peak position using both a Gaussian function and a Lorentzian function, to capture the elastic and inelastic contributions to the scattering, respectively. The field dependence of the Lorentzian contribution of the scattering for a (b) [1-10] magnetic field, (c) [001] magnetic field, and (d) [111] magnetic field. The field dependence of the Lorentzian contribution tracks the intensity of the low energy quasi Goldstone-mode excitations.}
\label{Lorentz}
\end{figure}

As discussed in the manuscript, elastic cuts of the (220) magnetic Bragg peak have been carefully modeled with a two component fit for all field strengths and orientations. An example of one such fit is shown in Fig.~\ref{Lorentz}(a) for the [111] sample at 0~T. The Gaussian component of the scattering originates from the nuclear and magnetic long-range order and its field dependence is thoroughly discussed in the manuscript. The second component of the scattering at (220) is captured by a Lorentzian function, and accounts for the inelastic part of the scattering due to the partial integration over the quasi Goldstone-mode excitations. The Lorentzian contribution of the scattering is somewhat interesting as it provides information on the very low energy inelastic field dependence. The field evolution of the integrated intensity of the Lorentzian component is shown for all field directions in Fig.~\ref{Lorentz}(b,c,d).\ 

For the [1-10] field direction, the Lorentzian part of the scattering reaches zero intensity for a field of 0.5~T (Fig.~\ref{Lorentz}(b)). This decrease of scattering is consistent with the complete removal of the quasi Goldstone-mode excitations (Fig.~\ref{Eslice}(a)). The Lorentzian contribution of the scattering maintains zero intensity up to 1.5~T, at which point a clear and abrupt increase is observed. This originates from the softening of the spin wave excitations towards the elastic line (Fig.~\ref{Eslice}(a) Fig.~\ref{Ecuts}(a)) and, thus, are picked up by our integration over the elastic channel. Above 1.5~T, no Lorentzian contribution of the scattering is observed in a [1-10] field.\  

For the [001] field direction, the Lorentzian contribution increases dramatically at 0.1~T, due to the enhancement of the scattering from the quasi Goldstone-mode excitations (Fig.~\ref{Lorentz}(c)). Indeed, as seen in Fig.~\ref{Eslice}(b) and Fig.~\ref{Ecuts}(b), the quasi Goldstone-mode excitations are clearly enhanced upon application of fields up to 0.5~T. Referring once again to Fig.~\ref{Lorentz}(c), we can see that above 0.1~T, the inelastic (Lorentzian) contribution steadily decreases, reaching zero intensity for fields above 0.75~T. This effect can also be explained by examining the energy cuts of Fig.~\ref{Ecuts}(b). Comparing the data sets between 0.1~T and 1~T, we see that the inelastic intensity shifts to progressively higher energies with increasing field. As we are integrating over our elastic resolution in Fig.~\ref{Lorentz}(c), from $-0.1$ to 0.1~meV, this integration picks up less of the inelastic contribution at higher fields as the intensity moves out of our elastic window. This shifting of the inelastic intensity to higher energies signals that the spin wave gap of the quasi Goldstone-mode is growing from 0.1~T up to 1~T. Above 1~T, the Lorentzian part of the scattering completely disappears as the (220) magnetic Bragg peak has zero intensity in the field polarized state.\ 

Lastly, we turn our attention to the field evolution of the Lorentzian component in a [111] field which is shown in Fig.~\ref{Lorentz}(d). Similar to the low field dependence of the elastic scattering (Fig.~\ref{Elascuts001}(f)), we observe that the inelastic intensity is flat up to 0.15~T. Above 0.15~T, the intensity falls off abruptly up to 1~T. This decrease of scattering is consistent with the reduced intensity of the quasi Goldstone-mode excitations (Fig.~\ref{Eslice}(c) and Fig.~\ref{Ecuts}(c)). As was the case for the [1-10] field, for 1.5~T, we observe a large enhancement of the Lorentzian intensity. This originate from the softening of the spin wave excitations towards the elastic line. This feature is best observed by looking at the 1.5~T energy slice shown in Fig.~\ref{Eslice}(c). Above 1.5~T, the spin wave excitations are pushed to higher energy and the Lorentzian contribution of the scattering remains at zero intensity.\

\end{appendix}

\bibliography{ETO.bib}

\end{document}